\documentclass[aps,prb,twocolumn,amsmath,amssymb,floatfix]{revtex4}
\usepackage{color}
\usepackage{amssymb}
\usepackage{wasysym}
\usepackage{graphicx}
\usepackage{dcolumn}
\usepackage{bm}

\newcommand{\ybco}{YBa$_2$Cu$_3$O$_{y}$}

\newcommand{\ybcoc}{YBa$_2$Cu$_3$O$_{6.45}$}

\newcommand{\ybcof}{YBa$_2$Cu$_4$O$_{8}$}

\newcommand{\lbcoo}{La$_{15/8}$Ba$_{1/8}$CuO$_4$}
\newcommand{\lsco}{La$_{2-x}$Sr$_x$CuO$_4$}
\newcommand{\lesco}{La$_{2-x-y}$Eu$_y$Sr$_x$CuO$_4$}

\newcommand{\bscco}{Bi$_2$Sr$_2$CaCu$_2$O$_{8+\delta}$}
\newcommand{\ccoc}{Ca$_{2-x}$Na$_x$CuO$_2$Cl$_2$}

\newcommand{\Tc}{T_{\rm c}}

\newcommand{\bk}{{\vec k}}
\newcommand{\bQ}{{\vec Q}}

\newcommand{\Qch}{{\vec Q}_{\rm c}}
\newcommand{\Qchx}{{\vec Q}_{{\rm c}x}}
\newcommand{\Qchy}{{\vec Q}_{{\rm c}y}}

\newcommand{\Qspx}{{\vec Q}_{{\rm s}x}}
\newcommand{\Qspy}{{\vec Q}_{{\rm s}y}}

\newcommand{\w}{\omega}

\begin{document}

\title{
Stripes and electronic quasiparticles\\ in the pseudogap state of cuprate superconductors
}

\author{Matthias Vojta}
\affiliation{Institut f\"ur Theoretische Physik, Technische Universit\"at Dresden,
01062 Dresden, Germany}

\date{\today}


\begin{abstract}
This article is devoted to a discussion of stripe and electron-nematic order and their
connection to electronic properties in the pseudogap regime of copper-oxide
superconductors. We review basic properties of these symmetry-breaking ordering phenomena
as well as proposals which connect them to quantum-oscillation measurements. Experimental
data indicate that these orders are unlikely to be the cause of the pseudogap phenomenon,
implying that they occur on top of the pseudogap state which itself is of different
origin. Specifically, we discuss the idea that the non-superconducting pseudogap ground
state hosts electron-like quasiparticles which coexist with a spin liquid, realizing a
variant of a fractionalized Fermi liquid. We speculate on how stripe order in such a
pseudogap state might offer a consistent description of ARPES, NMR, quantum-oscillation,
and transport data.
\end{abstract}

\maketitle


\section{Introduction}
\label{intro}

The field of cuprate high-temperature superconductors has seen remarkable experimental
progress in recent years. However, many conceptual puzzles remain in interpreting the
data, in particular when it comes to the non-superconducting state. One such puzzle
concerns the character and origin of the so-called pseudogap regime in underdoped
cuprates.\cite{timusk,norman_rev,np_feat} Numerous proposals have been made to explain this apparent
suppression of low-energy states below a doping-dependent pseudogap temperature $T^\ast$.
These proposals include phase-incoherent Cooper pairing, other orders competing with
superconductivity, exotic fractionalized states, and short-range singlet correlations as
precursor to the half-filled Mott insulator.\cite{norman_rev,leermp} Experiments continue
to shape and modfiy the community's view on the pseudogap regime, with some of the
central recent results being the observation of quantum oscillations at low temperature
and high magnetic field\cite{norman10,qosc_rev1,qosc_rev2} and the detection of
various symmetry-breaking non-superconducting orders.

Among these orders, so-called stripe order\cite{pnas,stripe_rev1,antonio_rev,stripe_rev2}
takes a prominent role. Static spin and charge density modulations have not only been
established\cite{jt95,jt96,abbamonte05} to exist in the \lsco\ (or 214) family of
cuprates, but they have also been detected in other families: Modulations in the charge
sector, likely pinned by impurities, appear on the surface of \ccoc, \bscco\ and other
Bi-based cuprates -- as seen by scanning-tunnelling-microscopy (STM)
techniques\cite{kapi03a,yazdani04,kohsaka07,wise08,parker10} -- and charge stripes have also
been found in \ybco\ in high magnetic fields.\cite{julien11} Electron-nematic order,
i.e., a breaking of lattice rotational symmetry, is often discussed in the context of
stripes, as the melting of uni-directional modulations can naturally lead to a state with
restored translation, but broken rotation symmetry.\cite{KFE98} Indeed, for \ybco\ it has
been argued that strong low-energy anisotropies in various
experiments\cite{hinkov08,ando02,taill10b} originate from the tendency toward electron-nematic
order which enhances the weak orthorhombic crystalline
anisotropy.\cite{yamase06,gros06,hackl09b} Finally, there is by now accumulated
evidence\cite{fauque06,fauque08a,fauque08b} for a certain type of loop-current order
which breaks lattice inversion but preserves translation symmetry.\cite{varma99}

The fact that all these observations occurred in the pseudogap regime prompts to
interrogate the relation between these symmetry-breaking orders and the pseudogap. In
fact, recent theoretical proposals attempt to explain quantum-oscillation and (a subset
of) transport data in the pseudogap regime in terms of Fermi-liquid-like quasiparticles
moving in spin and charge-modulated
states.\cite{millis07,sudip,linmillis,harrison09,millis10,hackl10,kiv11,harrison11,sudip_rev}
However, a fully coherent picture has not yet emerged. On a broader scale, it is
questionable that stripes account for the full pseudogap phenomenology. Alternatively, if
the pseudogap state is of different origin, stripes should be investigated {\em on top}
of this pseudogap state.

The purpose of this paper is twofold.
First, we review salient properties of stripe and associated nematic states, and
critically discuss current ideas on the emergence of Fermi pockets and their connection
to quantum oscillations.
Second, we discuss specific ideas\cite{pwa87,YRZ,YRZ_rev,qiss10,moonss11} for the pseudogap regime
based on holes moving in a spin liquid with short-range magnetic order and connect them
to the concept of fractionalized Fermi liquids.\cite{flst1,flst2} We then propose to
model stripes utilizing a phenomenological form of the single-particle propagator in the
pseudogap regime and analyze their properties.

In this article, the discussion will be mainly restricted to the non-superconducting
state, and ideas on the pairing mechanism will not be covered. Similarly, loop-current
order,\cite{varma99,ddw} albeit very interesting, will only be mentioned briefly.

\subsection{Outline}

The body of this paper is organized as follows:
In Sec.~\ref{sec:stripe} we introduce the stripe and related electron-nematic orders.
Sec.~\ref{sec:pockets} contains a general discussion on aspects of the cuprate pseudogap,
together with current ideas on how stripe order on top of a Fermi-liquid-like state might
explain experimental results from quantum-oscillation and transport measurements in the
pseudogap regime.
Conceptual problems encountered in this description will lead us in Sec.~\ref{sec:flst}
to consider a different state underlying the pseudogap, namely a state
where a spin-liquid background coexists with fermionic quasiparticles forming small Fermi
pockets. This realizes a variant of a fractionalized Fermi liquid.
Finally, Sec.~\ref{sec:yrzstr} contains concrete calculations for stripes in a pseudogap
state which itself is described by a phenomenological single-particle propagator
accounting for the formation of hole pockets. We discuss to
what extent these calculations can explain experimental observations.
An outlook will close the paper.


\section{Stripe and electron-nematic orders}
\label{sec:stripe}

Lattice-symmetry-breaking order in cuprates has been the subject of numerous review
articles in the past.\cite{pnas,stripe_rev1,antonio_rev,stripe_rev2} Here we concentrate
on density-wave (or stripe) order -- which breaks lattice translation symmetry and, in
its uni-directional form, lattice rotation symmetry -- and on so-called
electron-nematic order, which breaks rotation but not translation symmetry.
(In this liquid-crystal terminology, uni-directional density-wave order may
also be called electron-smectic order.)

In this section we shall review both the relevant order parameters and their
manifestation in various experimental probes.\cite{stripe_rev1,stripe_rev2,ssrmp}

\subsection{Order parameters and symmetry breaking}

Following Landau, ordered phases are best described in terms of order-parameter fields.
In the following, we assume that the underlying crystal lattice is two-dimensional, with
square-lattice (i.e. tetragonal) symmetry.

A charge density wave (CDW) is described by a pair of complex scalar fields
$\phi_{cx}$, $\phi_{cy}$ for the two CDW directions with wavevector $\Qchx$ and $\Qchy$.
The charge density (more generally, any observable which transforms as a scalar under
spin rotations) is assumed to obey
\begin{equation}
\langle \rho ({\vec R}, \tau) \rangle = \rho_{\rm avg} + \mbox{Re} \left[e^{i \Qch
\cdot {\vec R}} \phi_c (\vec R, \tau)  \right] \,.
\label{chargemod}
\end{equation}
Similarly, a collinear spin-density wave (SDW) requires a pair of complex vector fields
$\phi_{s\alpha x}$, $\phi_{s\alpha y}$, $\alpha=x,y,z$, with wavevectors $\Qspx$ and
$\Qspy$, such that the spin density follows
\begin{equation}
\langle S_\alpha ( {\vec R},\tau ) \rangle = \mbox{Re} \bigl[ e^{i{\vec Q}_s\cdot {\vec R}} \phi_{s\alpha} ({\vec R},\tau )
\bigr].
\label{sdw}
\end{equation}
Experimental results in 214 cuprates are consistent with uni-directional order, i.e.,
$\phi_{cx},\phi_{s\alpha x}\neq 0$ and $\phi_{cy}=\phi_{s\alpha y}=0$ or vice versa. The
preferred wavevectors are $\Qspx = 2\pi(0.5\pm 1/M,0.5)$, $\Qspy = 2\pi(0.5,0.5\pm 1/M)$
and $\Qchx = (2\pi/N,0)$, $\Qchy = (0,2\pi/N)$, where $M$ and $N$ are the
doping-dependent real-space periodicities which follow $M=2N$ to a good
accuracy.\cite{stripe_rev1,antonio_rev,stripe_rev2,jt95}

Given the propensity for uni-directional order, introducing a separate order parameter
for rotational symmetry breaking in a tetragonal environment is useful:\cite{KFE98}  this
is an Ising scalar $\phi_n$ for $l=2$ spin-symmetric electron-nematic order which carries
wavevector $\vec Q=0$. $\phi_n$ may be defined from any spin-singlet observable
which is even under time reversal and sensitive to real-space directions,
like the bond kinetic energy,
\begin{equation}
\phi_n(\vec r,\tau) =
\langle c_\sigma^\dagger(\vec r,\tau) c_\sigma(\vec r+x,\tau) - c_\sigma^\dagger(\vec r,\tau) c_\sigma(\vec
r+y,\tau)\rangle\,.
\end{equation}
It is common practice to refer to $\phi_n$ as nematic order
parameter; note, however, that $\phi_n\neq 0$ in both nematic and smectic (stripe)
phases. As discussed in Ref.~\onlinecite{KFE98}, a nematic phase can occur as a precursor
to stripe oder, but nematic order unrelated to stripes has been discussed for the
cuprates as well.\cite{yamase06}
In some cuprates the tetragonal in-plane symmetry is broken down to orthorhombic, and
spontaneous nematic order cannot exist. It may then still make sense to discuss the
tendency toward electron-nematic order if a small microscopic anisotropy becomes strongly
enhanced by electronic correlations.

Any of the above orders may co-exist with superconductivity, and arguments for both
competition\cite{zhang97,vs99,demler01,stripe_rev2} and
cooperation\cite{castellani96,zaanen01,inhom_sc03} of stripes and superconductivity have
been put forward.
While the anticorrelation between strong stripe order and high superconducting $\Tc$ and
the field-induced enhancement of stripe order in superconducting compounds point towards
competition,\cite{stripe_rev2} the so-called antiphase superconductivity
proposed\cite{berg07} to explain the intriguing behavior of \lbcoo\
(Ref.~\onlinecite{li07}) may point towards a more complex interplay of stripes and
pairing.\cite{pdw}

\subsection{Experimental signatures}

A variety of experimental probes may be used to detect stripe and electron-nematic
orders. Here we provide a brief list on theoretical expectations and corresponding
experimental observations.\cite{stripe_rev2}

Perhaps most importantly, broken translation symmetry arising from long-range
density-wave order results in additional superlattice Bragg reflections in diffraction
experiments. Those have been seen in 214 cuprates, both in the charge sector (using
neutrons and X-rays) and in the spin sector (using neutrons).\cite{jt95,jt96,abbamonte05}
Static long-range charge order has not been seen in other compounds; weak incommensurate
spin order is detected in very underdoped \ybco\ as well.\cite{haug10} Static magnetic
order has also been detected using $\mu$SR in both 214 and \ybco\ compounds, in both
cases broadly consistent with the results from neutron scattering.

Spatial inhomogeneities, i.e., site differentiation in local quantities, have been
detected using NMR and NQR, again in 214 and \ybco\ cuprates.\cite{stripe_rev2} While
most of this data could not be used to clearly infer an ordering pattern, a very recent
NMR experiment\cite{julien11} showed clear signatures of period-4 charge stripes in
\ybco\ exposed to large magnetic fields. Furthermore, stripe-like inhomogeneities in the
charge sector are seen in STM experiments on the surface of \bscco\ and
\ccoc.\cite{kapi03a,yazdani04,kohsaka07,parker10}

Broken translation symmetry also results -- via Bragg scattering -- in a reconstruction
of the dispersion relation of any elementary excitation in the solid, and should
therefore be visible, e.g., in single-electron and phonon spectra.
Unfortunately, the experimental situation is not clear-cut, i.e., the backfolding of
dispersions due to stripe order has not been unambigously observed. This suggests that those
effects might be weak and moreover masked by matrix-element effects.
We note, however, that ARPES in
stripe-ordered \lesco\ has found some hints for reconstructed electron
bands\cite{borisenko08} (recall that the quality of ARPES data in 214 cuprates does at present not
reach that of Bi-based cuprates). Also, various phonon anomalies have been detected and
discussed in the context of stripe order.\cite{stripe_rev2,reznik}

Finally, sizeable stripe order can lead to a modulation of collective magnetism, in the
extreme rendering it quasi-one-dimensional. The spin excitations of stripe-ordered
\lbcoo\ (Ref.~\onlinecite{jt04}) indeed display an elevated-energy dispersion akin to
that of spin ladders, and consequently models of coupled-spin ladders\cite{mvtu04,gsu04}
(or field-theoretic versions thereof\cite{vvk}) have been used to model the experimental
data.
Qualitatively similar results have also been obtained in a description of static
stripes using the time-dependent Gutzwiller approximation to the one-band Hubbard
model.\cite{seibold}

The broken rotation symmetry inherent to both uni-directional stripe and electron-nematic
states will cause in-plane anisotropies of all direction or momentum-dependent
properties, such as charge and heat transport, fluctuation spectra etc. However, these
anisotropies will only be visible in macroscopic measurements in a mono-domain situation.
The alternating layer distortions in the LTT phase of 214 compounds therefore preclude
global anisotropies. In \ybco\ this is different, and temperature-dependent anisotropies
both in transport coefficients\cite{ando02,taill10b} and in magnetic fluctuation
spectra\cite{hinkov08} have been observed, supporting the idea that the small (global)
crystalline anisotropy becomes enhanced by electron-correlation effects at low
temperatures. We note that the expected orthorhombic Fermi-surface distortion has not
been seen, but this may again be related to the rather low quality of ARPES data for
\ybco.

The anisotropic magnetic fluctuations discovered in neutron scattering on \ybcoc\ in the
absence of static order\cite{hinkov08} admit interpretations in terms of both genuine
(stripe-free) nematic order\cite{yamase09,lawler10} and incipient (i.e. fluctuating)
stripe order.\cite{vvk,vojta10}


\section{Stripes, Fermi pockets, and the pseudogap}
\label{sec:pockets}

In this section, we discuss existing ideas on how stripe order could explain
selected features of the pseudogap regime, in particular quantum-oscillation and transport
data. We shall argue that the existing modelling is likely incomplete because
stripe order alone cannot account for the full pseudogap.

\subsection{Pseudogap in a nutshell}

Experimental data characterizing the pseudogap regime of cuprate superconductors have
seen a spectacular evolution over the last years.\cite{timusk,norman10} To set the
stage, we start by summarizing the most important observations.

Most generally, the term ``pseudogap'' refers to a partial suppression of low-energy
electronic states, as compared to what is expected from a metallic phase,  upon cooling
below a doping-dependent pseudogap temperature $T^\ast$. This suppression is seen in both
thermodynamic and spectroscopic observables.
Momentum-resolved measurements indicate that the electronic states in the so-called
antinodal region near momenta $(0,\pm\pi)$ and $(\pm\pi,0)$ are the ones which become
strongly gapped, such that the Fermi surface is reduced to apparent ``Fermi
arcs''.\cite{kanigel}
Based on recent photoemission experiments it has been argued\cite{pdj11} that these arcs
in fact represent the ``front part'' of Fermi pockets, whose area is possibly compatible with
the hole doping level, and we will come back to this later.

Both $T^\ast$ and the estimated size of the low-temperature antinodal pseudogap,
$\Delta_{\rm PG}$, monotonically decrease with doping, in striking difference to the
superconducting $\Tc$. $T^\ast$ has been suggested to extrapolate to the scale $J$ of the
magnetic exchange in the limit of zero doping, and to vanish either around optimal doping
or at the overdoped end of the superconducting dome. It should be noted, however, that
the definition of $T^\ast$ is not unique but depends -- to some extend -- on the
considered observable.

In underdoped cuprates, signatures of short-range antiferromagnetism arising from local
moments are ubiquitous: This includes magnetic collective modes with large spectral
weight\cite{gu09,letacon11} and magnetism which is induced or enhanced by non-magnetic impurities
(e.g. by small concentrations of Zn substituting for
Cu).\cite{tallon99,tallon00,alloul_rev}
Conceptually, it is plausible to attribute these strong antiferromagnetic fluctuations to
the proximity to the parent Mott-insulating state,\cite{leermp} where local moments and
their spin-wave excitations determine the magnetic response.

Moreover, symmetry-breaking orders are frequently found in the pseudogap regime. Most
prominent are stripes\cite{pnas,stripe_rev1,antonio_rev,stripe_rev2} seen as long-range
bulk order in 214 cuprates and as disorder-pinned order on the surface of, e.g., \bscco\
and \ccoc. The tendency towards stripe order can be enhanced by applying a magnetic
field, such that stripe order can even be induced in compounds where zero-field static
order is absent.\cite{chang09,julien11} The most plausible interpretation of this effect
is based on competing order parameters: The magnetic field acts to suppress
superconductivity which in turn enhances competing order such as stripes.\cite{demler01}
In addition to stripes, strong temperature-dependent global anisotropies signifying
electron-nematic order have been detected in \ybco.\cite{stripe_rev1,stripe_rev2}
Lastly, a weak intra-unit-cell magnetic order has been detected in polarized neutron
scattering in both \ybco\ and
HgBa$_2$CuO$_{4+\delta}$,\cite{fauque06,fauque08a,fauque08b} which has been attributed to
circulating-current order of the type proposed by Varma.\cite{varma99}
It is remarkable that these various competing orders have been observed essentially
exclusively in the pseudogap regime, with interpretations to be discussed in
Sec.~\ref{sec:pgscen} below.

\subsection{Quantum oscillations}
\label{sec:osc}

Quantum oscillations, i.e., oscillations of thermodynamic and transport properties as
function of $1/B$ where $B$ is a large applied magnetic field, are known from standard
metals where they are most easily understood in terms of a semiclassical quantization of
cyclotron orbits. de Haas-van Alphen and Shubnikov-de Haas effects are described by the
Lifshitz-Kosevich theory.\cite{LK} At $T=0$, the observed oscillation frequency $F$ can
be related to the momentum-space area $A$ enclosed by the orbit at the Fermi energy via
\begin{equation}
\label{area}
F = \frac{\hbar c}{2\pi e} A \,.
\end{equation}

To date, quantum oscillations have been detected not only in overdoped
Tl$_2$Ba$_2$CuO$_{6+\delta}$ (Ref.~\onlinecite{vignolle08}) -- located outside the
pseudogap regime and expected to resemble a conventional metal -- but also in underdoped
\ybco\ and \ybcof,\cite{doiron07,yelland08,sebastian08} i.e., well inside the pseudogap regime.
Quantum oscillations have also been observed on the electron-doped side of the phase
diagram,\cite{gross1,gross2} where the data have been found to be broadly consistent with a Fermi surface
reconstructed by antiferromagnetic order with $\vec{Q}=(\pi,\pi)$.

The observation of quantum oscillations in weakly hole-doped cuprates is considered an
important breakthrough in the field, because it was long believed that coherent
electronic quasiparticles do not exist in the pseudogap regime -- an absence that was
hypothesized mainly because early photoemission experiments only detected broad features
in the single-electron spectrum.
The natural conclusion from the observation of quantum oscillations is the presence of
coherent fermionic quasiparticles moving (in a semiclassical picture) along closed
momentum-space orbits. In fact, experimental data are in favor of Fermi-liquid-like
quasiparticles causing the oscillations.\cite{sebastian10,ramshaw10}
Accepting this interpretation, various proposals for ``exotic''
pseudogap phases (e.g. with full spin--charge separation or otherwise incoherent
carriers) appear to be ruled out.

For the later discussion, a few details of the quantum-oscillation results need to be
mentioned:\cite{qosc_rev1,qosc_rev2}
\begin{itemize}

\item[(i)]
The observations are so far restricted to the YBCO family and a relatively narrow
doping range, $0.09 < x < 0.125$ in \ybco\ and $x\approx0.14$ in \ybcof. It is uncertain
whether the disappearance of quantum oscillations outside this doping range indicates a
distinct quantum phase transition (QPT)\cite{sebastian10b} or simply originates from lower sample
quality.

\item[(ii)]
Converting the measured oscillation frequency of approximately 530\,T into a
momentum-space area using Eq.~\eqref{area}, one concludes that the carriers form Fermi
pockets whose size is approximately 2\% of the Brillouin zone. This pocket size increases
with doping $x$, but slower than expected for a pocket area $\propto x$, see e.g. Fig.~5b
of Ref.~\onlinecite{qosc_rev1}.

\item[(iii)]
Transport measurements detect a negative Hall constant in the regime where quantum
oscillations occur, more precisely, in \ybco\ at large fields and low temperatures for
all $x>0.08$.\cite{leboeuf07,taill11} In a Fermi-liquid-based picture, this suggests that
quantum oscillations are caused by electron (instead of hole) pockets. (However, it has
been argued from experimental data that electron-like transport may also arise from
near-nodal states.\cite{jt_osc})

\item[(iv)]
There is no consensus on the role of superconductivity in the quantum-oscillation
experiments. Although the measurements are performed in a resistive state, pairing e.g.
in a ``vortex liquid'' can still be strong. While most interpretations assume that the
oscillations reflect the physics of the underlying normal state, it is also conceivable
that a superconducting state features Fermi pockets\cite{pdw,varma_priv,kallin12} which lead to quantum
oscillations.

\end{itemize}

\subsection{Fermi pockets from density waves}
\label{sec:flpockets}

The small momentum-space area of the Fermi pockets deduced from quantum-oscillation data
in underdoped cuprates [see point (ii) in Sec.~\ref{sec:osc} above] has triggered
explanations based on translation-symmetry breaking in a metallic normal state:
Density-wave order reconstructs the underlying Fermi surface, such that multiple pockets
can appear, whose individual is unrelated to the electron concentration (apart from an
overall sum rule).
Given that stripes are the most prominent form of density-wave order observed in the
cuprates, it is plausible to connect Fermi pockets caused by stripe order (in the spin or
charge channel) to quantum oscillations.

Concrete calculations are essentially all based on a mean-field description of the
symmetry-breaking order on top of a conventional Fermi-liquid-like state. Most
often\cite{millis07,sudip,linmillis,harrison09,millis10,hackl10,kiv11,harrison11} the
latter is simply modelled by free fermions with a dispersion $\epsilon_\bk$ and a
``large'' Fermi surface as derived from band-structure calculations. Then, the
quasiparticle properties of the ordered state can be obtained from the diagonalization of
a simple Hamiltonian matrix. For a uni-directional stripe state with $M=2N=8$ this matrix
takes the form
\begin{widetext}
\begin{equation}
\left[
\begin{array}{cccccccc}
 \varepsilon_\bk & V_c^\ast & 0 & V_c & 0 & V_s^\ast & V_s & 0\\
 V_c & \varepsilon_{\bk+(\frac{\pi}{2},0)} & V_c^\ast & 0 & 0 & 0 & V_s^\ast & V_s\\
 0 & V_c & \varepsilon_{\bk+(\pi,0)} & V_c^\ast & V_s & 0 & 0 & V_s^\ast\\
 V_c^\ast & 0 & V_c & \varepsilon_{\bk+(\frac{3\pi}{2},0)} & V_s^\ast & V_s & 0 & 0\\
 0 & 0 & V_s^\ast & V_s & \varepsilon_{\bk+(\frac{\pi}{4},\pi)} & V_c^\ast & 0 & V_c\\
 V_s & 0 & 0 & V_s^\ast & V_c & \varepsilon_{\bk+(\frac{3\pi}{4},\pi)} & V_c^\ast & 0\\
 V_s^\ast & V_s & 0 & 0 & 0 & V_c & \varepsilon_{\bk+(\frac{5\pi}{4},\pi)} & V_c^\ast\\
 0 & V_s^\ast & V_s & 0 & V_c^\ast & 0 & V_c & \varepsilon_{\bk+(\frac{7\pi}{4},\pi)}\\
\end{array}
\right] \ .
\label{mat}
\end{equation}
\end{widetext}
Here, $V_c$ ($V_s$) are the scattering potentials implementing long-range charge (spin)
stripe order. In general, $V_c$ and $V_s$ are complex and momentum-dependent, reflecting
the real-space structure and symmetry of the
order.\cite{millis07,linmillis,mvor08,hackl10} However, in most cases the Fermi surface
depends only weakly on the precise form of $V_c$ and $V_s$, which are therefore often
approximated as real constants.

\begin{figure*}[tb]
\begin{center}
\includegraphics[width=5in]{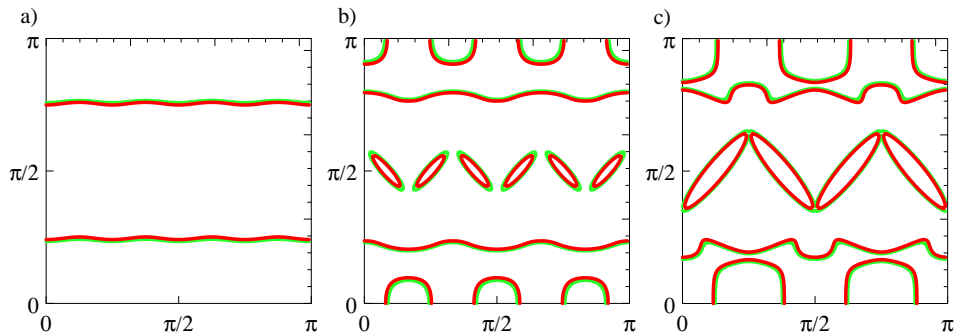}
\caption{
Doping evolution of Fermi surfaces (dark red) for vertical spin stripes, shown in a
quadrant of the Brillouin zone.
The stripe strength decreases with doping:
a) $x=1/16$, $M=16$, $V_s=0.04$\,eV.
b) $x=1/12$, $M=12$, $V_s=0.03$\,eV.
c) $x=1/8$, $M=8$, $V_s=0.02$\,eV.
In all cases, weak charge order is
added with $V_c=V_s/10$ and $N=M/2$. The light green lines are constant-energy contours
at small negative energy to indicate the dispersion gradient.
From b) to a) a Lifshitz transition occurs where the antinodal electron pockets disappear
(the near-nodal hole pockets disappear as well).
For further details see text.
}
\label{fig:fs1}
\end{center}
\end{figure*}

\begin{figure*}[bt]
\begin{center}
\includegraphics[width=5in]{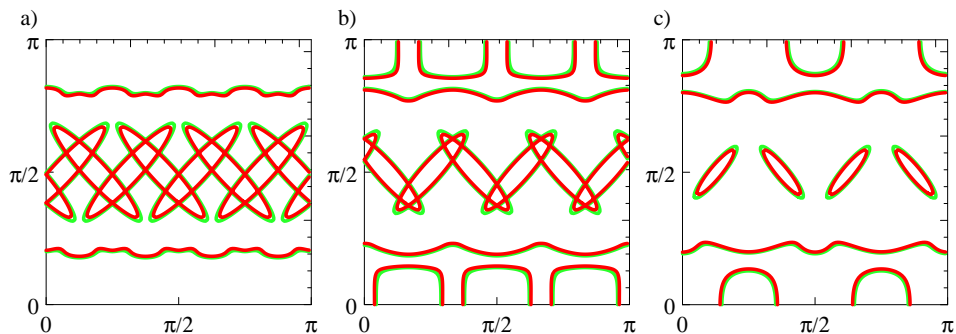}
\caption{
Same as Fig.~\ref{fig:fs1}, but now the stripes are strongest at $x=1/8$:
a) $x=1/16$, $M=16$, $V_s=0.015$\,eV. b) $x=1/12$, $M=12$,
$V_s=0.02$\,eV. c) $x=1/8$, $M=8$, $V_s=0.03$\,eV.
The near-nodal hole pockets exist for all doping levels.
}
\label{fig:fs2}
\end{center}
\end{figure*}

Uni-directional spin stripes have been found to produce electron pockets in the
reconstructed Fermi surface\cite{linmillis,millis10} which are formed out of antinodal electronic states.
These electron pockets have been proposed as explanation for the simultaneous occurance
of quantum oscillations and a negative low-temperature Hall coefficient in \ybco\ over a
certain range of dopings.\cite{taill11}
In this scenario, reducing the doping level may eventually lead to a Lifshitz transition
where the electron pockets merge and disappear in favor of open orbits.\cite{millis10}

Sample results, which are qualitatively similar to those published in
Refs.~\onlinecite{millis07,linmillis,hackl10,millis10}, are shown in Figs.~\ref{fig:fs1}
and \ref{fig:fs2}. To facilitate comparison to the results in Sec.~\ref{sec:yrzstr}
below, the dispersion is taken from renormalized mean-field theory,\cite{YRZ}
$\epsilon_\bk = - 2t (\cos k_x + \cos k_y) - 4 t' \cos k_x \cos k_y - 2 t''(\cos 2k_x +
\cos 2k_y)$ with $t_0=0.3$\,eV, $t_0'=-0.3t_0$, $t_0''=0.2t_0$ and renormalized hopping
amplitudes $t=g_t(x) t_0 + 3 g_s(x) J \chi / 8$, $t'=g_t(x) t_0'$, $t''=g_t(x) t_0''$,
where $\chi=0.338$ and the Gutzwiller factors are $g_t(x) = 2x/(1+x)$, $g_s(x)=4/(1+x)^2$.

Comparing Figs.~\ref{fig:fs1} and \ref{fig:fs2} it is obvious that the area of all Fermi
pockets is determined by an interplay of doping level, stripe period, and stripe strength
and thus depends on microscopic details. The disappearance of the antinodal electron
pockets happens due to the doping dependence of the spatial modulation period, which is
assumed here to be similar to that of 214 cuprates.\cite{yamada98}
It should be noted that, in addition to electron pockets, near-nodal hole pockets as in
Figs.~\ref{fig:fs1} and \ref{fig:fs2} occur quite generically in such stripe models. It
has been suggested that the signatures of these pockets in quantum-oscillation and Hall
measurements might be weak due to a low carrier mobility, but we are not aware of a
satisfactory explanation of the latter. A related problem may be that high-field
specific-heat measurements\cite{boeb11} find a rather small value of $\gamma=C/T$, which
appears incompatible with multiple Fermi sheets as in Figs.~\ref{fig:fs1}b,c and
\ref{fig:fs2}b,c.

Based on some analogies between 214 and \ybco\ cuprates, it has been suggested that the
Fermi-surface reconstruction by stripe order is a general phenomenon, being responsible
for a number of thermodynamic and transport anomalies.\cite{taill09,taill11b} While
stripe order indeed appears to be realized in both cuprate families, we believe that care
is required when it comes to ideas about ``fluctuating stripe order''. Given that d.c.
transport and quantum oscillations are essentially static probes, they will {\em not} be
influenced by temporally fluctuating stripe order. In other words, only static order with
sufficiently large correlation length can be expected to leave signatures in transport
coefficients. (This is different for a finite-energy probe like inelastic neutron
scattering which is sensitive to fluctuating stripes.\cite{vvk})

A few alternative proposals for density-wave order deserve to be mentioned, which were
partially motivated by (a) the apparent lack of strong spin
order\cite{hinkov08,haug10,julien11} in \ybco\ for $0.09< x < 0.14$ and (b) the problem of
invoking low-energy antinodal states for the pockets which, however, are believed to be
strongly gapped in the pseudogap regime.
Ref.~\onlinecite{kiv11} combined charge-only stripe order with a sizeable nematic distortion of
the Fermi surface and obtained electron pockets for $x=1/8$, however, an account of the
doping dependence was not given.
In contrast, Ref.~\onlinecite{harrison11} proposed bi-directional charge order which may
yield electron pockets formed out of nodal electron states. It remains open how this
proposal can be compatible with the results of scattering and NMR experiments.

We will return to this set of ideas and their conceptual aspects in Sec.~\ref{sec:concprob}
below.

\subsection{Scenarios for the pseudogap}
\label{sec:pgscen}

The presentation so far prompts to ask for the relation between stripe order and the
pseudogap phenomenon, urging to discuss the origin of the latter. Given that the list of
proposed explanations for the pseudogap is long, I will restrict this section to a
summary of the most influential suggestions.

One class of ideas describes the pseudogap as a precursor of superconductivity, i.e., the
suppression of fermionic DOS happens because of phase-incoherent Cooper-pair formation.
Indeed, signatures of preformed pairs above $\Tc$ \cite{preformed} have been identified
in a number of experiments, most notably Nernst effect measurements,\cite{nernst_exp1}
but also in photoemission,\cite{valla08,kanigel08} and STM studies.\cite{yazdani07}
This interpretation is supported by the observation of fluctuating diamagnetism which
often varies in proportion to the Nernst coefficient.\cite{nernst_diamag} However, the
experimental onset temperature of pairing fluctuations is significantly below the
$T^\ast$ established from other probes. This casts strong doubts on the assumption of
preformed pairs being the primary source of the pseudogap.

A second class ascribes the pseudogap to an ordering phenomenon which competes with
superconductivity. The vanishing of the corresponding ordering temperature upon
increasing the doping level naturally induces a quantum critical point (QCP) on the
doping axis, which could be made responsible for anomalous normal-state properties. Among
the concrete proposals for such competing phases are stripe\cite{rome95,castellani98} and
electron-nematic\cite{KFE98} phases as well as circulating-current
orders.\cite{varma99,ddw}
One objection against these proposals is that the pseudogap line at $T^\ast$ does not
appear to be associated with a thermodynamic phase transition. However, this can be
circumvented either by invoking quenched disorder which tends to smear the transition or
by postulating a special form of phase transition with weak thermodynamic singularities,
e.g. of Kosterlitz-Thouless or Ashkin-Teller type\cite{sudbo08b}. In fact, a weak but
sharp signature in the uniform susceptibility has recently been detected in YBCO samples
of different doping, tracking the pseudogap temperature.\cite{monod08}
A second objection concerns the momentum-space structure of the gap: a finite-$\bQ$
order parameter (like stripes) causes gaps to appear near certain hot spots, but this is
inconsistent with the apparent $d$-wave nature of the pseudogap.\cite{timusk}
A third objection, perhaps most serious, concerns universality: The pseudogap appears to
be a universal phenomenon in hole-doped cuprates, with very similar properties in the
different families, which suggests a common origin. In contrast, stripes are strongest in
single-layer compounds of the 214 family, but are weak or absent, e.g., in materials with
more than two CuO$_2$ layers per unit cell.  A
similar objection may also apply to loop-current order at $\bQ\!=\!0$, with their
signatures detected in a subset of cuprate families only.
Finally, while microscopic calculations using cluster extensions of dynamical mean-field
theory (DMFT), to be discussed in more detail in Sec.~\ref{sec:cdmft} below, have
established the existence of a pseudogap in the 2d Hubbard
model,\cite{macridin06,stanescu06,jarrell08,civelli08,imada08} this pseudogap occurs in
the {\em absence} of long-range order. This suggests that long-range order is not
required for pseudogap formation.

A conceptual remark is in order: A particular symmetry-breaking order being the origin of the pseudogap
obviously explains why this order is only seen below $T^\ast$, while discarding
such a pseudogap explanation would imply that the various symmetry-breaking orders are a
result (rather than the cause) of the pseudogap.
(A recent detailed analysis of STM data from \bscco\ concluded that stripes are unlikely
to be the cause of the pseudogap.\cite{parker10})

We are lead to a third class of ideas, relating the pseudogap to strong correlations and
the proximity to the parent Mott insulator, without invoking symmetry breaking. While a
satisfactory phenomenological description of these ideas is lacking to date, the
theoretical results can be grouped into (i) numerical results from cluster extensions of DMFT, which suggest
that short-range singlet formation may be responsible for a partial gap
formation,\cite{macridin06,stanescu06,jarrell08,civelli08,imada08}, (ii) corresponding field
theories trying to formalize ``Mottness'',\cite{phillips} and (iii) proposals based on
low-energy theories for exotic phases involving fractionalization. Such proposals also
imply the existence of a QCP on the doping axis (marking a transition between the
underdoped exotic and the overdoped Fermi-liquid phase), but here {\em without} symmetry
breaking on either side of the phase diagram.
One concrete such proposal will be the subject of Sec.~\ref{sec:flst} below.

\subsection{Conceptual problems of Fermi-liquid based stripes}
\label{sec:concprob}

Clearly, identifying the origin of the pseudogap also has immediate consequences for the
proposed explanations for quantum-oscillation experiments. A key question is whether a
conventional Fermi-liquid-based description of the symmetry-breaking orders is justified.
While the quantum-oscillation data itself suggest that the answer might be ``yes'', the
temperature evolution of both thermodynamic and spectral properties in the pseudogap
regime\cite{timusk} appears inconsistent with this assumption.

A generally accepted solution of this puzzle is not known to date, but two alternatives
are obvious:

\begin{itemize}

\item[(A)]
The low-doping state is asymptotically a conventional Fermi liquid (in the absence of
superconductivity), but with both coherence temperature and quasiparticle weight being
small. This implies, e.g., that the underlying Fermi surface of this state is ``large'',
i.e., it fulfills Luttinger's theorem (see Sec.~\ref{sec:flst} for a more detailed
discussion). This large Fermi surface may be unobservable if translational symmetry
breaking sets in above the coherence temperature.

\item[(b)]
The low-doping state is a metallic non-Fermi liquid which features coherent fermionic
quasiparticles, but is {\em not} a Fermi liquid because Luttinger's theorem is violated.
Such a state can be classified as fractionalized Fermi liquid,\cite{flst1} discussed in
detail in Sec.~\ref{sec:flst}. Most plausibly, the Fermi surface consists of hole
pockets,\cite{qiss10,moonss11} with a total area given by the doping level $x$ -- this
appears supported by cluster DMFT studies.\cite{imada08}
A phenomenological ansatz for the self-energy describing a pseudogap state with small
pockets has been put forward independently in Ref.~\onlinecite{YRZ}.

\end{itemize}

To date, photoemission experiments have not been able to distinguish the available
scenarios, mainly because of insufficient energy and momentum resolution
(although progress has been made recently\cite{fournier10,pdj11}).

Most theoretical models to explain the quantum-oscillation frequencies in underdoped
cuprates operate with density-wave order occurring on top of a Fermi-liquid state with
weakly interacting quasiparticles and a large Fermi surface. Apparently, these models are
heavily based on scenario (A).
Taking the models of Refs.~\onlinecite{millis07,millis10} and others at face value, a
number of problems are apparent:
\begin{itemize}

\item[(1)] The single-particle states which form the electron pockets stem from the antinodal
portion of the Brillouin zone where, however -- according to ARPES -- a large gap exists.
While this argument may be flawed because it is comparing the high-field
low-temperature regime of quantum oscillations with the zero-field elevated-temperature
regime of ARPES, the more general problem is that a stripe modulation on top of a large
Fermi surface does not account for the pseudogap (e.g. because it cannot explain the
momentum-space structure of the pseudogap).

\item[(2)] Stripe order leads, in addition to the desired electron pockets, to
further sheets of Fermi surface (both open orbits and hole pockets). Their existence
appears to be in conflict with high-field specific-heat data: The small measured value of
$\gamma$ suggests that the pockets seen in quantum oscillations reflect the only gapless
charge carriers in the system.\cite{boeb11}

\item[(3)] In most calculations, sizable spin stripe order is required to produce electron
pockets while charge order alone leads to hole pockets only. However, a recent NMR
experiment on \ybco\ suggests that spin order is not present (or very weak) in the regime
of interest.\cite{julien11} Charge stripes on top of a nematic state have been
suggested to resolve this conflict.\cite{kiv11}

\item[(4)] The weakly doping-independent area of the pocket in the doping range
$0.09<x<0.14$ for \ybco\ and \ybcof requires fine tuning in essentially all models,
perhaps with the exception of the bi-directional charge modulation proposed
in Ref.~\onlinecite{harrison11}.

\end{itemize}

Clearly, the conceptual problem (1) appears most pressing. This prompts to consider
scenario (B), i.e., a non-Fermi liquid pseudogap state, on top of which stripe order may
occur. This is discussed in the remainder of the article.


\section{Fractionalized Fermi liquids and the pseudogap}
\label{sec:flst}

We now turn to an appealing scenario for the pseudogap ground state, which invokes a
non-Fermi liquid which nevertheless features well-defined fermionic quasiparticles. Such
a state is best classified as a ``fractionalized Fermi liquid'' (FL$^\ast$). This term,
originally introduced in the context of heavy-fermion systems, refers to a metallic phase
without spontaneously broken symmetries in which sharp quasiparticles (with spin 1/2 and
charge $e$) coexist with a paramagnetic spin liquid.\cite{osmott_rev}

We begin by reviewing the concepts of fractionalized Fermi liquids and the related
orbital-selective Mott transitions for two-band systems (like heavy fermions) and then
discuss the application of these ideas to weakly doped (one-band) Mott insulators. This
will eventually lead us to propose the phase diagram in Fig.~\ref{fig:pd2}, where a
fractionalized Fermi liquid at low doping and a conventional Fermi liquid at high doping
are separated by a QPT.

\subsection{Orbital-selective Mott phases and fractionalized Fermi liquids in two-band systems}

Loosely speaking, an orbital-selective (or band-selective) Mott phase -- dubbed OS Mott
in the follwing -- is a metallic phase of a two-band system, where one band (dubbed $c$
in the following) is ``metallic'' (with Fermi-liquid-like quasiparticles), while the
other ($f$) is ``Mott-insulating'' as a result of strong electronic correlations. (The
generalization to more than two bands is straightforward.) The electrons in the
Mott-insulating band form local moments, which may either order magnetically or settle in
a non-magnetic spin-liquid state at low temperatures. As will be argued below, a
spin-liquid ground state of the local moments is required to define a fractionalized
Fermi liquid.

A sharp distinction between a band being metallic or
Mott-insulating can be made by considering the momentum-space volume enclosed by the
Fermi surface: here a metallic band contributes, whereas a Mott-insulating band does not.
In a standard Fermi-liquid (FL) phase of a two-band system, both bands are metallic in the
above sense, and the Fermi volume is given by
\begin{equation}
{\cal V}_{\rm FL} = K_d (n_{\rm tot}\,{\rm mod}\,2).
\label{vfl}
\end{equation}
Here $n_{\rm tot}=n_c+n_f$ denotes the number of electrons per unit cell, ${\rm mod}\,2$
implies that full bands are not counted, $K_d = (2\pi)^d/(2 v_0)$ is a phase space
factor, with $v_0$ the unit cell volume, and the factor of 2 accounts for the spin
degeneracy of the bands. Importantly, the ``large'' Fermi volume of Eq.~\eqref{vfl} is in
accordance with Luttinger's theorem.\cite{oshi2}

The key point is now that an OS Mott phase has a Fermi volume {\em different} from ${\cal
V}_{\rm FL}$ \eqref{vfl} because the electrons of the Mott-insulating band do not
contribute. As the number of electrons {\em per site} corresponding to a Mott-insulating
band is unity, $n_{\rm tot}$ in Eq.~\eqref{vfl} is replaced by $(n_{\rm tot}\!-\!N_{\rm
cell})$, which yields a distinct Fermi volume if the number of sites per unit cell,
$N_{\rm cell}$, is odd (and spin degeneracy is preserved).
Therefore, a sharp definition of an OS Mott phase requires that the local moments of the
Mott-insulating band are in a spin-liquid state without spontaneously broken symmetries
-- such a state typically features fractionalized spin excitations. Hence, we arrive at a
phase where conduction electrons coexist with a fractionalized spin liquid -- this is
exactly what was dubbed ``fractionalized Fermi liquid'' in Ref.~\onlinecite{flst1}. Its
Fermi volume is ``small'',
\begin{equation}
{\cal V}_{\rm FL^\ast} = K_d [(n_{\rm tot}-1)\,{\rm mod}\,2],
\label{vflst}
\end{equation}
and violates the Fermi-volume count according to Luttinger's theorem exactly by unity,
hence this FL$^\ast$ phase is a true non-Fermi liquid metal. It is separated from FL by a
quantum phase transition dubbed ``orbital-selective Mott transition'', where the Fermi
volume changes discontinuously.

If, in contrast, the local moments in the OS Mott regime realize a symmetry-broken state
as $T\to 0$, be it an antiferromagnet or a valence-bond solid, the unit cell is enlarged,
and the sharp distinction between small and large Fermi volumes is lost (because the two
states only differ in the number of {\em full} bands). Consequently, the resulting
local-moment phase may be adiabatically connected to a fully itinerant broken-symmetry
Fermi liquid.\cite{mv_af}

FL$^\ast$ phases can be readily constructed in the framework of Kondo-lattice
models.\cite{flst1,flst2}
Start from a spin-only system with single-site unit cell and geometric frustration, such
that a spin-liquid state is realized. Candidates are spin-1/2 magnets on the Kagome or
pyrochlore lattices as well as e.g. triangular or square lattices with longer-range or
multiple spin-exchange interactions.\cite{schmidt10,white11,balents11} Then, add
conduction electrons and couple those via a (weak) Kondo coupling to the spin liquid. As
the spin liquid is a stable state of matter, a weak Kondo coupling is irrelevant in the
RG sense, hence the properties of the two subsystems will remain qualitatively unchanged,
and a FL$^\ast$ phase naturally emerges. (This is particularly obvious if the spin liquid
is gapped.)
Hence, in the Kondo limit the existence of FL$^\ast$ phases in two-band systems only
rests upon the existence of fractionalized spin liquids in frustrated magnets.
In this limit $n_f\equiv 1$, and ${\cal V}_{\rm FL^\ast}$ \eqref{vflst} is given by $n_c$
only. The OS Mott transition between FL$^\ast$ and FL can be understood as breakdown of
Kondo screening of the local moments which happens due to the competition between Kondo
effect and non-local inter-moment correlations.\cite{coleman01,si01,flst1,pepin07}
(In transition-metal oxides realizing multiband Hubbard systems, an additional driver for
OS Mott physics is Hund's-rule coupling which counteracts the screening of local
moments.\cite{osmott_rev,medici05}

We note that, in principle, different flavors of FL$^\ast$ phases can exist, depending on
the nature of the spin-liquid component (gapped, gapless, $Z_2$ or $U(1)$ gauge forces).
However, the full systematics of FL$^\ast$ phases, including the fate of topological
order in the presence of charge carriers, is not understood to date.

\subsection{Application to weakly doped Mott insulators}
\label{sec:flst1b}

Various phenomenological aspects of underdoped cuprates suggest an interpretation in
terms of a fractionalized Fermi liquid, where hole-like charge carriers coexist with a
spin-liquid background. This thinking goes back to Anderson's early idea of a resonating
valence-bond state doped with holes\cite{pwa87,KRS87} and has been subsequently reformulated in
various flavors.\cite{poil96,YRZ,YRZ_rev,kaul08,qiss10,moonss11,wen11}

\begin{figure}
\begin{center}
\includegraphics[width=3in]{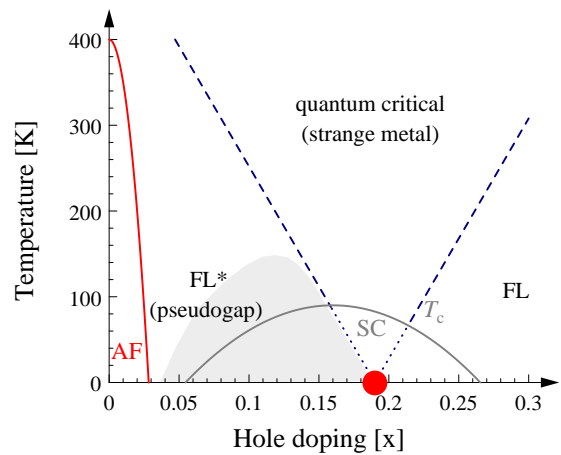}
\caption{
Hypothesized phase diagram for hole-doped cuprates in the temperature--doping plane, with
a quantum critical point (${\color{red}\CIRCLE}$) separating the overdoped FL and the
underdoped FL$^\ast$ phase -- the latter represents the pseudogap, which itself may host
various forms of competing order (shaded). The finite-temperature quantum critical
regime of the FL--FL$^\ast$ transition is cut-off below $T_c$ by superconductivity.
Also shown is the onset of commensurate antiferromagnetism (AF); not shown are additional
crossovers associated with the onset of pairing fluctuations and with spin-glass behavior.
}
\label{fig:pd2}
\end{center}
\end{figure}

A number of conceptual ideas are naturally account for:
(i) local moments inherited from the parent Mott insulator continue to exist in the weakly
hole-doped FL$^\ast$ state,
(ii) the charge-carrier density in this weakly doped Mott insulator is small (i.e. given
by the number of doped holes instead of the total number of electrons), and these
carriers may form small hole pockets,
(iii) the low-doping phase cannot be adiabatically connected to a Fermi liquid with a large Fermi
surface, but instead a QPT occurs on the doping axis, which in turn may be responsible
for extended non-Fermi-liquid behavior around optimal doping above $\Tc$,
(iv) electronic states in certain momentum-space regions (where only the large Fermi
surface would exist) will disappear at low energy and temperature in the low-doping
FL$^\ast$ phase, thereby creating a pseudogap, and
(v) a superconducting state formed at low doping will display a pairing gap on the
small Fermi surface of FL$^\ast$ which coexists with the pseudogap.
Apparently, this list bears striking similarity with experimental observations in cuprates.

Given the precise definition of a fractionalized Fermi-liquid phase in a two-band system,
it is natural to ask how this concept can be adapted to a one-band doped Mott
insulator.\cite{3band_foot}
It is plausible to start again from an undoped Mott insulator, where magnetic order is
suppressed by quantum fluctuations or frustration, such that a fractionalized spin liquid
is realized. A possible candidate state is an RVB-like state with short-range
correlations.
In contrast to the two-band case, charge carriers are now added in the {\em same} band by
doping holes. {\em If} these holes retain their integrity, i.e., continue to exist as
elementary excitations of spin 1/2 and charge $e$ with a Fermi surface, then an FL$^\ast$
phase is be realized: These quasiparticles co-exist with a spin liquid, and the
volume of the {\em hole} Fermi surface is given by the hole concentration $x$, whereas
Luttinger's theorem would dictate a Fermi volume of $(1-x)$ electrons or, equivalently,
$(1+x)$ holes. Hence, the Luttinger count is violated exactly by unity as in the two-band
situation discussed above.\cite{zero_foot}

A conceptual problem of this idea is that carrier doping of a fractionalized spin liquid
will -- quite generically -- lead to a fractionalization of the doped carriers,
i.e., a doped hole will decay into a neutral spinon (as this is a good quasiparticle of
the background spin liquid) and a charged spinless holon.\cite{KRS87,wen91,z2}
The key idea to circumvent this problem is via the formation of {\em bound
states}.
Imagine a fractionalized state in which a spinon and a holon can form
a low-energy bound state. This bound state is equivalent to a hole, and particles
occupying these bound states will feature a Fermi surface, with properties as described
above. As a spin-off, spinon-spinon bound states are possible as well, leading to sharp
low-energy resonances in neutron scattering. Importantly, such an FL$^\ast$ state will
display bound-state holes and fractionalization at the same time, such that low-energy
observables will often see signatures of sharp quasiparticles while experiments at higher
energy will detect continua characteristic of fractionalization.

Interestingly, the idea of spinons and holons undergoing ``reconfinement'' to form a
conventional quasiparticle was already discussed in the 1990s based on numerical results
for the $t-J$ model.\cite{poil96}
A phenomenological ansatz for a Green's function describing quasiparticles with hole
pockets was proposed by Yang, Rice, and Zhang;\cite{YRZ} its form can be related to
bound-state formation of holons and spinons in a mean-field RVB spin liquid.
More recently, Sachdev and co-workers\cite{qiss10,moonss11} studied the dynamics of
electrons in a fluctuating 2d antiferromagnet. This first leads to the emergence of an
algebraic charge liquid,\cite{kaul08} with separate excitations for spin and charge. Those were
proposed to form bound states, i.e., electrons with a pocket-like Fermi surface. A clear
connection to FL$^\ast$ phases was then made in Ref.~\onlinecite{moonss11}. In this
scenario, the spin-liquid component is gapless and derived from local AF order, instead
of the short-range RVB state advocated above.
Finally, Ref.~\onlinecite{wen11} also proposed a metallic state with hole pockets for the
pseudogap, which was dubbed a Luttinger-volume-violating Fermi liquid. It derives again
from a RVB-like spin-liquid state, where the doped hole retain their integrity. In line
with the above discussion, this state can be labelled fractionalized Fermi liquid as well.

\subsection{Numerical evidence}
\label{sec:cdmft}

Computer simulations of the 2d Hubbard model
\cite{haule07,jarrell08,civelli08,imada08,werner09,ferrero09,gull10} using cluster extensions of
DMFT\cite{cdmft_rev}
have provided insight into the physics of weakly doped Mott insulators. Among other
things, these calculations show pseudogap behavior at low doping in the absence of
superconductivity or magnetic order.

In the single-particle Green's function, this pseudogap behavior is accompanied by the
disappearance of the Fermi surface in parts of the Brillouin zone, namely in the
antinodal regions near $(\pi,0)$, $(0,\pi)$
\cite{civelli08,imada08,werner09,ferrero09,gull10}. This low-doping behavior is found to
be qualitatively distinct from that at high doping, with a regime of strong scattering in
between, suggestive of a quantum phase transition.\cite{haule07,jarrell08} These results
bear remarkable resemblance to those from photoemission experiments on actual cuprates.
Because of the similarity to orbital-selective Mott physics, the partial disappearance of
the Fermi surface upon decreasing the doping has been termed ``momentum-selective Mott
transition''.

This lends support to a phase diagram as in Fig.~\ref{fig:pd2}, where a non-Fermi liquid
phase is realized at low doping, which is separated by a QPT from a Fermi liquid at large
doping. In fact, a detailed analysis and extrapolation of the cluster-DMFT self-energies and
spectral functions in the low-doping regime has uncovered that the underlying Fermi
surface consists of the proposed four pockets.\cite{imada08} However, the detailed nature of the
accompanying spin-liquid state has not been understood in phenomenological terms to date.

\subsection{Single-particle propagator in the pseudogap phase}

While a complete theory for a one-band FL$^\ast$ is not available at present, concrete
proposals have been made for approximate forms of the single-electron Green's function in
the putative zero-temperature normal state describing the pseudogap. These include the
phenomenlogical ansatz by Yang, Rice, and Zhang (YRZ, Ref.~\onlinecite{YRZ}) and the
effective model of Qi and Sachdev (QS, Ref.~\onlinecite{qiss10}) for bound-state
formation in an algebraic charge liquid.

\begin{figure*}[tb]
\begin{center}
\includegraphics[width=5in]{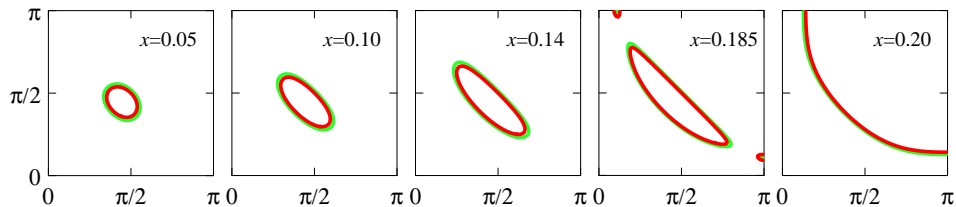}
\caption{
Fermi surfaces in the pseudogap state as described by the YRZ ansatz (\ref{yrz},\ref{yrz2}) for
different doping levels. For $x\lesssim0.18$ only the lower YRZ band crosses the Fermi
level, leading to one hole pocket per quadrant of the Brillouin zone. For $x>x_c=0.2$ the
large Fermi surface is recovered.
}
\label{fig:fs_yrz}
\end{center}
\end{figure*}

Although the underlying physical ideas are somewhat different, both proposals can be
written in terms of the following single-particle propagator
\begin{equation}
G(\bk,\w) = \frac{Z}{\w-\xi_\bk-\Delta_\bk^2/(\w-\xi'_\bk)}
\label{yrz}
\end{equation}
defined on the microscopic square lattice. Here $\xi_\bk$ is a single-particle dispersion
leading to a large Fermi surface, such that the large-doping FL state is described by
$\Delta_\bk=0$, and $Z$ is a (doping-dependent) weight factor.
In the YRZ proposal, the individual terms are given by
\begin{eqnarray}
\xi_\bk &=& \epsilon_\bk-\mu\,, \nonumber \\
\xi'_\bk &=& 2t (\cos k_x + \cos k_y)\,,\nonumber \\
\Delta_\bk &=& \Delta(x) (\cos k_x-\cos k_y)\,,
\label{yrz2}
\end{eqnarray}
with $\epsilon_\bk = - 2t (\cos k_x + \cos k_y) - 4 t' \cos k_x \cos k_y - 2 t''(\cos 2k_x + \cos
2k_y)$.
Physically, $\xi'_\bk$ reflects a nearest-neighbor ``spinon'' dispersion, while the
$d$-wave form of the gap $\Delta_\bk$ arises from an implicit pairing assumption.
In contrast, the QS scenario yields
\begin{eqnarray}
\xi_\bk &=& \epsilon_{1\bk}+\epsilon_{2\bk}-\mu\,, \nonumber \\
\xi'_\bk &=& \epsilon_{1,\bk+\bQ}-\epsilon_{2,\bk+\bQ}-\mu\,, \nonumber \\
\Delta_\bk &=& \Delta\,.
\end{eqnarray}
Here, $\Delta$ measures the strength of local antiferromagnetic order, and
$\epsilon_{1\bk}$ and $\epsilon_{2\bk}$ are the effective dispersions of the bound
states between the two species of spinless fermions and bosonic spinons which appear in
the algebraic charge liquid. Interestingly, this ``fermion doubling'' reflects the
topological order of the underlying spin liquid.\cite{qiss10}

For finite $\Delta_\bk$, Eq.~\eqref{yrz} leads to hole-pocket Fermi surfaces as shown in
Fig.~\ref{fig:fs_yrz}, with the exact shape and location being somewhat different in the
YRZ and QS approaches.
Both proposals require that, for $\Delta_\bk\neq0$, the {\em hole} area enclosed by the
Fermi pockets is given $x$ holes, i.e., the {\em electron} area is $(2-x)$, such that the
Fermi-volume count violates Luttinger's theorem by unity. In both cases, a line of zeroes
of $G$ emerges where $\xi'_\bk=0$ -- note that this coincides with the antiferromagnetic
zone boundary, $|k_x|+|k_y|=\pi$, only in the YRZ case.\cite{zero_foot}

In comparison to experiments, the Fermi surfaces derived from the YRZ ansatz have been
argued to be in good agreement with pockets deduced from recent ARPES
results,\cite{pdj11} and agreement with other thermodynamic and spectroscopic observables
has been suggested as well.\cite{YRZ_rev}
On the theoretical side, an analysis of the electronic self-energy obtained from
small-cluster diagonalization of the Hubbard model has found the YRZ ansatz to be a
reasonable approximation of the data.\cite{eder11}
Taken together, this suggests that an ansatz of the form \eqref{yrz} is an appropriate
starting point in order to model symmetry-breaking orders in the pseudogap regime.


\section{Stripes in the pseudogap regime}
\label{sec:yrzstr}

From the discussion above, it appears plausible to take a different route in modelling
electronic properties of stripes, namely to add stripe modulations on top of a
pseudogapped ground state. In the absence of a complete theory of the latter, it is
suggestive to combine the proposed forms of the single-particle propagator in a FL$^\ast$
state (YRZ or QS, see Sec.~\ref{sec:flst1b}) with a mean-field description of stripe order
-- this is what we will discuss in the following. To explore the connection to quantum
oscillations and transport, we exclusively concentrate on the Fermi surface of such a
stripe state.

To be specific, we employ the YRZ propagator, Eqs.~(\ref{yrz},\ref{yrz2}), which can be
written as a sum of two coherent quasiparticle poles at energies $E_\bk^\pm$, dubbed upper and lower
YRZ band. In the relevant doping range, the upper YRZ band lies far above the Fermi
level\cite{YRZ_rev} and can thus be ignored. Hence, we consider quasiparticles in the
lower YRZ band which undergo stripe order. (This is similar in spirit to the treatment of
superconductivity in the YRZ pseudogap state in Ref.~\onlinecite{yrzsc} which considered
pairing of the lower-band YRZ quasiparticles only.) At the mean-field level, stripe order is
captured by Bragg scattering terms which connect $\bk$ and $\bk+\bQ$ as usual, such that
we are lead to diagonalize a matrix as in Eq.~\eqref{mat}, where the bare energies are
replaced by $E_\bk^-$. Momentum-dependent weight factors of the YRZ quasiparticles are
ignored at this level, but those will only have a minor influence on the Fermi surface.

\begin{figure*}[tb]
\begin{center}
\includegraphics[width=4.5in]{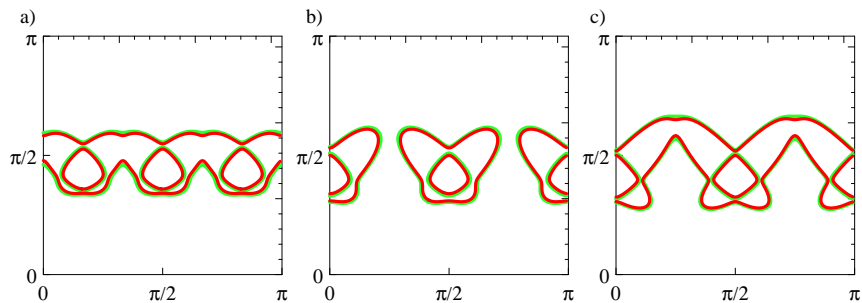}
\caption{
Fermi surfaces (as in Fig.~\ref{fig:fs1}), but here for vertical charge stripes in a
pseudogap state with hole pockets, the latter described by the YRZ ansatz.\cite{YRZ}
a) $x=1/12$, $N=6$, $V_c=0.04$\,eV.
b) $x=1/10$, $N=4$, $V_c=0.03$\,eV.
c) $x=1/8$, $N=4$, $V_c=0.025$\,eV.
For details see text.
}
\label{fig:fs11}
\end{center}
\end{figure*}

\begin{figure*}[tb]
\begin{center}
\includegraphics[width=4.5in]{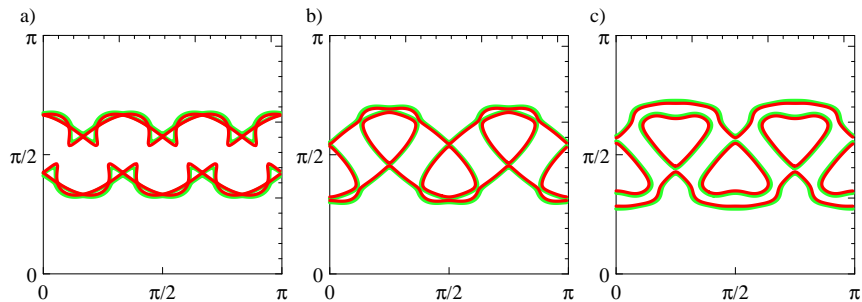}
\caption{
Fermi surfaces for pseudogap stripes as in Fig.~\ref{fig:fs11}, but now for vertical spin
stripes.
a) $x=1/12$, $M=12$, $V_s=0.035$\,eV.
b) $x=1/10$, $M=8$, $V_s=0.035$\,eV.
c) $x=1/8$, $M=8$, $V_s=0.05$\,eV.
In all cases, weak charge order is added with $V_c=V_s/10$ and $N=M/2$.
}
\label{fig:fs12}
\end{center}
\end{figure*}

We now turn to the results of this approach. We have chosen numerical parameters for the
dispersion as in Sec.~\ref{sec:flpockets}, and the YRZ hybridization is given by
$\Delta(x) = \Delta_0 (1-x/x_c)$ with $\Delta_0 = t_0/2$ and $x_c=0.2$.\cite{YRZ}
Sample Fermi surfaces are shown in Figs.~\ref{fig:fs11} and \ref{fig:fs12}, for the cases
without and with spin order, respectively.
The scattering potentials have been chosen to be moderate to strong, leading to relative
doping modulations of order unity, to better visualize hybridization effects and pocket
formation.

As is clearly visible, modulations on top of a hole-pocket state lead, depending on
microscopic parameters, to a Fermi-surface reconstruction into hole pockets of different
sizes and/or open orbits. The pockets are generically rather small and, by construction,
emerge from near-nodal states. It appears, however, essentially impossible to generate
electron pockets in this approach -- both with and without spin order -- because all
low-energy bands are weakly hole-doped. Also, most Fermi-surface deformations arise from
the hybridization between equal-spin bands, i.e., from charge order or from secondary
spin-order effects. Quantitatively, the area of the small pockets in Fig.~\ref{fig:fs11}
(Fig.~\ref{fig:fs12}c) ranges between 0.6\ldots1\% (1.6\%) of the full Brillouin zone, the
large pockets in Fig.~\ref{fig:fs11}b have a size of approximately 4\% of the full Brillouin
zone.

From these results we conclude that stripe order on top of a pseudogap state with hole
pockets is not easily consistent with the quantum oscillation data, because the resulting
Fermi pockets are either too large or too small in area. Of course, this conclusion
hinges on the applicability of the semiclassical analysis of the quantum oscillation
data, which is not guaranteed for a one-band FL$^\ast$ state. (For a two-band FL$^\ast$
with asymptotically decoupled bands, such an analysis can be expected to be ok.)
The fact that all pockets are hole-like also makes it difficult to explain the negative Hall
coefficient, but here physics beyond weakly interacting quasiparticles might be
relevant.\cite{jt_osc}

Alternatively, one could speculate about a more severe reconstruction of the FL$^\ast$
state by stripe formation, not captured by simply subjecting the YRZ quasiparticles to a
mean-field stripe potential. This requires a thorough discussion of the interplay between stripy
modulations and the driving forces of FL$^\ast$, and will be presented in a forthcoming
publication.


\section{Discussion}

In the first part of this paper, we have reviewed central aspects of stripe and nematic
order in the cuprate high-temperature superconductors and connected them to observable
properties in the pseudogap regime. In particular, we have critically discussed the
proposal of electron pockets formed by density-wave order in a large Fermi-surface
state which has been suggested to explain quantum oscillation and transport data.

Based on the fact that such descriptions do not account for full pseudogap physics -- in
particular, the large Fermi surface is not seen experimentally -- we have suggested in
the second part of this paper to investigate stripes on top of a pseudogap state. Here,
we have concentrated on states of the fractionalized-Fermi-liquid type where coherent
quasiparticles populating hole pockets coexist with a spin-liquid background. Employing
the YRZ description of the single-particle propagator, we have determined the Fermi
surfaces of such pseudogap stripe states. These states -- which now account for salient
aspects of the pseudogap -- can, however, not easily explain the quantum oscillation
data.

This leaves us with a dilemma: Neither modulated Fermi-liquid states nor modulated
pseudogap (i.e. FL$^\ast$) states seem to account for the phenomenology of both quantum
oscillations and the pseudogap. The following is a (certainly incomplete) list of escape
routes:

\begin{itemize}

\item
The pseudogap physics seen a zero field above $\Tc$ is entirely irrelevant for the
quantum-oscillation regime at high fields and very small $T$. Then, a Fermi-liquid-based
approach to quantum oscillations could be justified. This option would require to explain
how a magnetic field of 30\ldots50\,T -- still small on electronic scales -- can destroy
the pseudogap (which otherwise appears rather robust and is not affected\cite{jt_osc} by
fields of order 10\,T).

\item
The low-energy fermions of the pseudogap state undergo a severe reconstruction in a
magnetic field, e.g., due to stripe formation. (Stripe formation in the
quantum-oscillation regime of \ybco\ has been experimentally established using
NMR.\cite{julien11}) We note that the formation of bound states of the fractionalized
constituents of FL$^\ast$ is a low-energy phenomenon which could in principle be altered
by moderate fields or modulations.

\item
The quantum oscillations do not arise from Fermi-liquid-like carriers, such that the
standard Lifshitz-Kosevich analysis\cite{LK} is not applicable. This would suggest to search for
quantum oscillations in other candidate pseudogap states.\cite{tami}

\item
Superconducting pairing cannot be neglected in the discussion of quantum oscillations. As
conventional superconducting states (i.e. with a full gap or with discrete nodes) are not
expected to cause quantum oscillations, one is lead to look for paired states with a
Fermi surface, which appears e.g. if the condensate carries a finite
momentum.\cite{pdw,varma_priv} The question of the character of the normal-conducting
pseudogap state may then be secondary.

\end{itemize}

Clearly, more experimental information would be helpful to make progress: seeing quantum
oscillations in other underdoped cuprates and/or over a larger doping regime would be
extremely interesting, as well as a further characterization of the stripe order in the
quantum-oscillation regime (i.e. doping/field/temperature dependence of the strength and
the modulation period). For theory, understanding the pseudogap -- and stripe order on
top of it -- remains the most important issue. Among other things, it should be
clarified whether and how a negative Hall coefficient may arise in hole-pocket (or
Fermi-arc) states.


\begin{acknowledgments}
The author acknowledges useful discussions with A.~Hackl, C. Proust, S. Sachdev,
L.~Taillefer, C. M. Varma, and A.~Wollny. This research has been supported by the DFG
through FOR 960 and GRK 1621.
\end{acknowledgments}



\begin{thebibliography}{99}
\bibitem{3band_foot}
Although a complete low-energy description of cuprates requires three bands (for the
copper and oxygen orbitals), it is believed that crucial aspects like pairing and the
pseudogap emerge already within a reduced one-band description.
\end{thebibliography}
\end{document}